\documentclass[prb,preprintnumbers,amsmath,amssymb,floatfix]{revtex4}

\usepackage{graphicx}
\usepackage{subfigure}
\usepackage{dcolumn}

\begin{document}

\title{Quantum effect of one-dimensional photonic crystal}
\author{Xiang-Yao Wu$^{a}$ \footnote{E-mail: wuxy2066@163.com},
 Ji Ma$^{a}$, Xiao-Jing Liu$^{a}$,
 Yu Liang$^{a}$ \\ Nuo Ba$^{a}$, Wan-Jin Chen$^{a}$\footnote{E-mail: chenwwjj@126.com}, Hong-Chun Yuan$^{b}$ and Heng-Mei
 Li$^{b}$}

\affiliation{a. Institute of Physics, Jilin Normal University,
Siping 136000 China\\\\ $^{b}$ {\small College of Optoelectronic
Engineering, Changzhou Institute of Technology, Changzhou 213002,
China}}
\begin{abstract}
In this paper, we have studied the quantum transmission
characteristics of one-dimensional photonic crystal with and
without defect layer by the quantum theory approach, and compared
the calculation results of classical with quantum theory. We have
found some new quantum effects in the one-dimensional photonic
crystal. When the incident angle $\theta=0$, there is no quantum
effect. When the incident angle $\theta\neq0$, we find there are
obvious quantum effect with the incident angle increase. At the
incident angle $\theta\neq0$, there are also quantum effect with
the change of thickness and refractive indexes of medium $B$ or
$A$. We think the new quantum effect of photonic crystal shall
help us to design optical devices.

\vskip 5pt

\end{abstract}

\maketitle {\bf 1. Introduction} \vskip 8pt

In 1987, E. Yablonovitch and S. John had pointed out that the
behavior of photons. It can be changed when propagating in the
material with periodical dielectric constant, and termed such
material Photonic Crystal [1, 2]. Photonic crystal important
characteristics are: Photon Band Gap, defect states, Light
Localization and so on. These characteristics make it able to
control photons, so it may be used to manufacture some high
performance devices which have completely new principles or can
not be manufactured before, such as high-efficiency semiconductor
lasers, right emitting diodes, wave guides, optical filters,
high-Q resonators, antennas, frequency-selective surface, optical
wave guides and sharp bends [3, 4], WDM-devices [5, 6], splitters
and combiners [7]. optical limiters and amplifiers [8-10]. The
research on photonic crystals will promote its application and
development on integrated photoelectron devices and optical
communication. To investigate the structure and characteristics of
band gap, there are many methods to analyze Photonic crystals
including the plane-wave expansion method [11], Green¡¯s function
method, finite-difference time-domain method [12-14] and transfer
matrix method [15-20]. All of methods are come from classical
Maxwell equations. In Refs. [21, 22], we have firstly studied the
the quantum transmission characteristics of one-dimensional
photonic crystal by the quantum theory approach, in which we have
only considered the incident angle is zero, i.e., vertical
incidence, we have found the classical and quantum transmission
characteristics are the completely same, i.e., there is not
quantum effect in one-dimensional photonic crystal. In this paper,
we have studied the quantum transmission characteristics of
one-dimensional photonic crystal when the incident angle is an
arbitrary angle, i.e., non-vertical incidence. we find there are
obvious quantum effect with the incident angle increase. At the
incident angle $\theta\neq 0$, there are also obvious quantum
effect with the change of thickness and refractive indexes of
medium $B$ or $A$. Otherwise, we have considered the effect of
defect layer on the quantum transmission characteristics. When the
incident angle $\theta=0$, there is also not quantum effect. When
the incident angle $\theta\neq0$, with the incident angle
increase, there are obvious quantum effect for the one-dimensional
photonic crystal with defect layer.

\vskip 8pt {\bf 2. The quantum transmissivity} \vskip 8pt

In Refs. [23, 24], with the quantum theory approach, we have
studied one-dimensional photonic crystal quantum transmissison
characteristic when the incident photon is vertical incidence,
i.e., the incident angle $\theta=0$. In the paper, we shall study
one-dimensional photonic crystal quantum transmissison
characteristic when the incident photon is non-vertical incidence,
i.e., the incident angle $\theta\neq0$.

The quantum wave equations of photon in medium is [23, 24]
\begin{equation}
\nabla\times\vec{\psi}=\frac{E-V}{c\hbar}\vec{\psi}=\frac{\omega}{c}n\vec{\psi},
\end{equation}
where $\omega$ is angle frequency of photon, $c$ is the velocity
of photon, $\vec{\psi}$ is the wave function of photon and $n$ is
refractive index of medium.

The incident light, reflected light and transmission light are in
the $xz$ plane, the incident angle is $\theta$, which are shown in
Fig. 1.
\begin{figure}[tbp]
\includegraphics[width=12 cm]{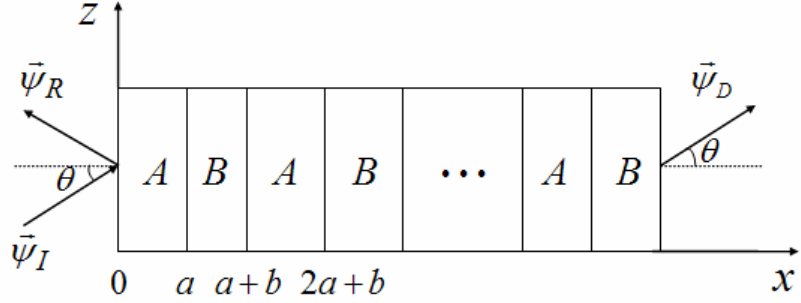}
\caption{The quantum structure of one-dimensional photonic
crystal.}
\end{figure}
The wave vectors $K_I$ and $K_R$ of incident and reflection photon
are
\begin{eqnarray}
\vec{K}_{I}=K_x\vec{i}+K_{z}\vec{k},\hspace{0.15in}
\vec{K}_{R}=-K_x\vec{i}+K_{z}\vec{k}
\end{eqnarray}
where $K_x=K\cos\theta$, $K_z=K\sin\theta$ are the wave vectors in
$x$ and $z$ direction, and $K=\frac{\omega}{c}$, the $\omega$ is
the incident photon angle frequency. The wave functions
$\vec{\psi}_{I}$ and $\vec{\psi}_{R}$ of incident and reflection
photon can be written as
\begin{equation}
\vec{\psi}_{I}=F_{x}e^{i(K_xx+K_{z}z)}\vec{i}+F_{y}e^{i(K_xx+K_zz)}\vec{j}+F_{z}e^{i(K_xx+K_zz)}\vec{k},
\end{equation}
\begin{equation}
\vec{\psi}_{R}=F'_{x}e^{i(-K_xx+K_zz)}\vec{i}+F'_{y}e^{i(-K_xx+K_zz)}\vec{j}+F'_{z}e^{i(-K_xx+K_zz)}\vec{k},
\end{equation}
where $F_{x}$ ($F'_{x}$), $F_{y}$ ($F'_{y}$) and $F_{z}$
($F'_{z}$) are the amplitudes of incident (reflected) photon wave
functions in the $x$, $y$ and $z$ directions.

Substituting $\vec{\psi}_{I}$ into Eq. (1), we have
\begin{eqnarray}
&&\left | \begin{array}{lll}
 \hspace{0.15in}\vec{i}   \hspace{0.9in}\vec{j}
 \hspace{0.85in}\vec{k}\\\\
 \hspace{0.1in}\frac{\partial}{\partial x}   \hspace{0.8in}\frac{\partial}{\partial y}   \hspace{0.75in}\frac{\partial}{\partial
 z}\\\\
 \hspace{0.0in}F_xe^{i(K_xx+K_zz)}   \hspace{0.1in}F_ye^{i(K_xx+K_zz)}
 \hspace{0.1in}F_ze^{i(K_xx+K_zz)}\\\\
 \end{array}
   \right |\nonumber\\&&=\frac{\omega}{c}(F_{x}e^{i(K_xx+K_zz)}\vec{i}+F_{y}e^{i(K_xx+K_zz)}\vec{j}+F_{z}e^{i(K_xx+K_zz)}\vec{k}),
   \end{eqnarray}
or
\begin{eqnarray}
\left \{\begin{array}{ll} \frac{\partial}{\partial
y}F_ze^{i(K_xx+K_zz)}-\frac{\partial}{\partial
z}F_ye^{i(K_xx+K_zz)}=\frac{\omega}{c}F_{x}e^{i(K_xx+K_zz)},\\\\
-\frac{\partial}{\partial
x}F_ze^{i(K_xx+K_zz)}+\frac{\partial}{\partial
z}F_xe^{i(K_xx+K_zz)}=\frac{\omega}{c}F_{y}e^{i(K_xx+K_zz)},\\\\
\frac{\partial}{\partial
x}F_ye^{i(K_xx+K_zz)}-\frac{\partial}{\partial
y}F_xe^{i(K_xx+K_zz)}=\frac{\omega}{c}F_{z}e^{i(K_xx+K_zz)},
    \end{array}
   \right.
\end{eqnarray}
i.e.,

\begin{eqnarray}
\left \{\begin{array}{ll}
-iK_zF_y=\frac{\omega}{c}F_x=KF_x, \\\\
-iK_xF_z+iK_zF_x=\frac{\omega}{c}F_y=KF_y, \\\\
iK_xF_y=\frac{\omega}{c}F_z=KF_z,
\end{array}
   \right.
\end{eqnarray}
From Eq. (7), we get
\begin{eqnarray}
F_x=-\frac{iK_z}{K}F_y,
\end{eqnarray}
\begin{eqnarray}
-iK_xF_z=(K-\frac{K^2_z}{K})F_y,
\end{eqnarray}
and
\begin{eqnarray}
K^2_x+K^2_z=K^2.
\end{eqnarray}
The incident current is [23, 24]
\begin{eqnarray}
\vec{J}_{I}&=&ic\vec{\psi}_{I}\times\vec{\psi}^*_{I}\nonumber\\
&=&2c[\frac{K^2-K^2_z}{KK_x}\vec{i}+\frac{K_z}{K}\vec{k}]\cdot|F_y|^2.
\end{eqnarray}
We find the incident current $\vec{J}$ is related to the amplitude
$F_y$ of $y$ component. So, we should only consider the $\vec{j}$
component wave function of photon in the following calculation.

Firstly, we study the transfer matrices in the first period. The
wave function of photon in medium $A$ is
\begin{eqnarray}
\vec{\psi}_{A}^1=[A_x^1e^{i(K_{Ax}x+K_{Az}z)}+A'^1_xe^{i(-K_{Ax}x+K_{Az}z)}]\vec{i} \nonumber\\
+[A_y^1e^{i(K_{Ax}x+K_{Az}z)}+A'^1_y e^{i(-K_{Ax}x+K_{Az}z)}]\vec{j} \nonumber\\
+[A_z^1 e^{i(K_{Ax}x+K_{Az}z)}+A'^1_z
e^{i(-K_{Ax}x+K_{Az}z)}]\vec{k},
\end{eqnarray}
where $A^1_{x}$ ($A'^1_{x}$), $A^1_{y}$ ($A'^1_{y}$) and $A^1_{z}$
($A'^1_{z}$) are the amplitudes of incident (reflected) photon
wave functions in the $x$, $y$ and $z$ directions.

Substituting $\vec{\psi}_{A}^1$ into Eq. (1), we have
\begin{equation}
-\frac{\partial}{\partial
z}(A^1_ye^{i(K_{Ax}x+K_{Az}z)}+A'^1_ye^{i(-K_{Ax}x+K_{Az}z)})=\frac{\omega}{c}n_A(A^1_xe^{i(K_{Ax}x+K_{Az}z)}+A'^1_{x}e^{i(-K_{Ax}x+K_{Az}z)}),
\end{equation}

\begin{eqnarray}
-\frac{\partial}{\partial x}(A^1_ze^{i(K_{Ax}x+K_{Az}z)}+A'^1_ze^{i(-K_{Ax}x+K_{Az}z)})+\frac{\partial}{\partial z}(A^1_xe^{i(K_{Ax}x+K_{Az}z)}+A'^1_xe^{i(-K_{Ax}x+K_{Az}z)}) \nonumber \\
\hspace
{3in}=\frac{\omega}{c}n_A[A^1_ye^{i(K_{Ax}x+K_{Az}z)}+A'^1_ye^{i(-K_{Ax}x+K_{Az}z)}],
\end{eqnarray}

\begin{eqnarray}
&&\frac{\partial}{\partial
x}(A^1_ye^{i(K_{Ax}x+K_{Az}z)}+A'^1_ye^{i(-K_{Ax}x+K_{Az}z)})
=\frac{\omega}{c}n_A[A^1_ze^{i(K_{Ax}x+K_{Az}z)}+A'^1_ze^{i(-K_{Ax}x+K_{Az}z)}].
\end{eqnarray}
From Eq. (13) to (15), we obtain
\begin{eqnarray}
\frac{A^1_y}{A'^1_y}=\frac{A^1_x}{A'^1_x}=-\frac{A^1_z}{A'^1_z}.
\end{eqnarray}
In the incidence zone, the total wave function is
\begin{eqnarray}
\vec{\psi}_{tot}(x,y,z)&=&\vec{\psi}_{I}(x,y,z)+\vec{\psi}_{R}(x,y,z)\nonumber\\
&=&(F_{x}e^{i(K_{x}x+K_{z}z)}+F'_{x}e^{i(-K_{x}x+K_{z}z)})\vec{i}\nonumber\\
&+&(F_{y}e^{i(K_{x}x+K_{z}z)}+F'_{y}e^{i(-K_{x}x+K_{z}z)})\vec{j}\nonumber\\
&+&(F_{z}e^{i(K_{x}x+K_{z}z)}+F'_{z}e^{i(-K_{x}x+K_{z}z)})\vec{k},
\end{eqnarray}
In the following, we should use the condition of wave function and
its derivative continuation at interface of two mediums.\\
(1) At $x=0$, by the continuation of wave functions
$\vec{\psi}_{tot}(x,y,z)$ and $\vec{\psi}_{A}^1(x,y,z)$, we have
\begin{eqnarray}
&&(F_{x}+F'_{x})e^{iK_{z}z}\vec{i}+(F_{y}+F'_{y})e^{iK_{z}z}\vec{j}
+(F_{z}+F'_{z})e^{iK_{z}z}\vec{k}\nonumber\\
&&=(A'^1_x+A'^1_{x})e^{iK_{z}z}\vec{i}+(A^1_{y}+A'^1_{y})e^{iK_{z}z}\vec{j}
+(A^1_{z}+A'^1_{z})e^{iK_{z}z}\vec{k},
\end{eqnarray}
(2) At $x=0$, by the derivative continuation of wave functions
$\vec{\psi}_{tot}(x,y,z)$ and $\vec{\psi}_{A}(x,y,z)$, we have
\begin{eqnarray}
&&iK_{x}(F_{x}-F'_{x})e^{iK_{z}z}\vec{i}+iK_{x}(F_{y}-F'_{y})e^{iK_{z}z}\vec{j}
\nonumber\\&&+iK_{x}(F_{z}-F'_{z})e^{iK_{z}z}\vec{k}\nonumber\\
&&=iK_{Ax}(A'^1_x-A'^1_{x})e^{iK_{Az}z}\vec{i}+iK_{Ax}(A^1_{y}-A'^1_{y})e^{iK_{Az}z}\vec{j}
\nonumber\\&&+iK_{Ax}(A^1_{z}-A'^1_{z})e^{iK_{Az}z}\vec{k},
\end{eqnarray}
with Eqs. (18) and (19), we get the $\vec{j}$ component relations
\begin{eqnarray}
(F_{y}+F'_{y})e^{iK_{z}z}=(A^1_{y}+A'^1_{y})e^{iK_{Az}z},
\end{eqnarray}
and
\begin{eqnarray}
K_{x}(F_{y}-F'_{y})e^{iK_{z}z}=K_{Ax}(A^1_{y}-A'^1_{y})e^{iK_{Az}z},
\end{eqnarray}
by Eqs. (20) and (21), we obtain
\begin{eqnarray}
K_z=K_{Az},
\end{eqnarray}
and
\begin{eqnarray}
\left \{\begin{array}{lll}
A^1_{y}+A'^1_{y}=F_{y}+F_{y}'\\
A^1_{y}-A'^1_{y}=\frac{K_{x}}{K_{Ax}}(F_{y}-F_{y}')
    \end{array}\right.,
\end{eqnarray}
by Eq. (23), we can obtain the matrix form of $A^1_{y}$,
$A'^1_{y}$ and $F_{y}$, $F_{y}'$
\begin{eqnarray}
\left ( \begin{array}{ll}
A^1_{y}\\
A'^1_{y}\\
   \end{array}
   \right )&&=\frac{1}{2} \left ( \begin{array}{ll}
1+\frac{K_{x}}{K_{Ax}}\hspace{0.2in} 1-\frac{{K_{x}}}{{K_{Ax}}}\\
1-\frac{{K_{x}}}{{K_{Ax}}} \hspace{0.2in} 1+\frac{{K_{x}}}{{K_{Ax}}}\\
   \end{array}
   \right )\left ( \begin{array}{ll}
F_{y}\\
F_{y}'\\
   \end{array}
   \right )\nonumber\\
   &&=M_{A}^{1}\left ( \begin{array}{ll}
F_{y}\\
F_{y}'
   \end{array}
   \right ),
\end{eqnarray}
where
\begin{eqnarray}
M_{A}^1&&=\frac{1}{2} \left ( \begin{array}{ll}
1+\frac{K_{x}}{K_{Ax}}\hspace{0.2in} 1-\frac{{K_{x}}}{{K_{Ax}}}\\
1-\frac{{K_{x}}}{{K_{Ax}}} \hspace{0.2in} 1+\frac{{K_{x}}}{{K_{Ax}}}\\
   \end{array}
   \right ),
\end{eqnarray}
is the transfer matrix of medium $A$ in the first period, and the
$K_{Ax}$ is
\begin{eqnarray}
K_{Ax}=\sqrt{K_{A}^{2}-K_{Az}^{2}}=\sqrt{K_{A}^{2}-K_{z}^{2}}=\sqrt{K_{A}^{2}-K^{2}\sin^{2}\theta},
\end{eqnarray}
where $K_{A}=\frac{\omega}{c}n_a$, $n_a$ is the refractive indexes
of medium $A$, and the transfer matrix $M_{A}^1$ can be written as
\begin{eqnarray}
M_{A}^1&&=\frac{1}{2} \left ( \begin{array}{ll}
1+\frac{K\cos\theta}{\sqrt{K_{A}^{2}-K^{2}\sin^{2}\theta}}\hspace{0.2in} 1-\frac{K\cos\theta}{\sqrt{K_{A}^{2}-K^{2}\sin^{2}\theta}}\\
1-\frac{K\cos\theta}{\sqrt{K_{A}^{2}-K^{2}\sin^{2}\theta}} \hspace{0.2in} 1+\frac{K\cos\theta}{\sqrt{K_{A}^{2}-K^{2}\sin^{2}\theta}}\\
   \end{array}
   \right ).
\end{eqnarray}
The wave function of photon in medium $B$ is
\begin{eqnarray}
\vec{\psi}_{B}^1(x,y,z)&=&(B^1_{x}e^{i(K_{Bx}x+K_{Bz}z)}+B'^1_{x}e^{i(-K_{Bx}x+K_{Bz}z)})\vec{i}\nonumber\\
&+&(B^1_{y}e^{i(K_{Bx}x+K_{Bz}z)}+B'^1_{y}e^{i(-K_{Bx}x+K_{Bz}z)})\vec{j}\nonumber\\
&+&(B^1_{z}e^{i(K_{Bx}x+K_{Bz}z)}+B'^1_{z}e^{i(-K_{Bx}x+K_{Bz}z)})\vec{k},
\end{eqnarray}
where $B^1_{x}$ ($B'^1_{x}$), $B^1_{y}$ ($B'^1_{y}$) and $B^1_{z}$
($B'^1_{z}$) are the amplitudes of incident (reflected) photon
wave functions in the $x$, $y$ and $z$ directions.

(3) At $x=a$, by the continuation of $\vec{j}$ component wave
functions $\vec{\psi}_{A}^1(x,y,z)$, $\vec{\psi}_{B}^1(x,y,z)$ and
their derivative, we have
\begin{eqnarray}
(A^1_{y}+A'^1_{y})e^{iK_{Az}z}+A^1_{y}e^{iK_{Ax}a}+A'^1_{y}e^{-iK_{Ax}a}
=(B^1_{y}+B'^1_{y})e^{iK_{Bz}z}+B^1_{y}e^{iK_{Bx}a}+B'^1_{y}e^{-iK_{Bx}a},
\end{eqnarray}
and
\begin{eqnarray}
iK_{Ax}(A^1_{y}e^{iK_{Ax}a+iK_{Az}z}-A'^1_{y}e^{-iK_{Ax}a+iK_{Az}z})
=iK_{Bx}(B^1_{y}e^{iK_{Bx}a+iK_{Bz}z}-B'^1_{y}e^{-iK_{Bx}a+iK_{Bz}z}),
\end{eqnarray}
with Eq. (29), we have
\begin{eqnarray}
(A^1_{y}+A'^1_{y})e^{iK_{Az}z}-(B^1_{y}+B'^1_{y})e^{iK_{Bz}z}=0,
\end{eqnarray}
\begin{eqnarray}
A^1_{y}e^{iK_{Az}a}+A'^1_{y}e^{-iK_{Az}a}-(B^1_{y}e^{iK_{Bx}a}+B'^1_{y}e^{-iK_{Bx}a})=0,
\end{eqnarray}
with Eq. (30), we get
\begin{eqnarray}
K_{Ax}(A^1_{y}-A'^1_{y})e^{iK_{Az}z}-K_{Bx}(B^1_{y}-B'^1_{y})e^{iK_{Bz}z}=0,
\end{eqnarray}
\begin{eqnarray}
K_{Ax}(A^1_{y}e^{iK_{Ax}a}-A'^1_{y}e^{-iK_{Ax}a})-K_{Bx}(B^1_{y}e^{iK_{Bx}a}+B'^1_{y}e^{-iK_{Bx}a})=0,
\end{eqnarray}
with Eqs. (31) and (33), we obtain
\begin{eqnarray}
K_{Az}=K_{Bz}.
\end{eqnarray}
The Eqs. (32) and (34) can be written as
\begin{eqnarray}
\left \{\begin{array}{ll}
A^1_ye^{iK_{Ax}a}+A'^1_ye^{-iK_{Ax}a}=B^1_ye^{iK_{Bx}a}+B'^1_ye^{-iK_{Bx}a}\\
K_{Ax}(A^1_ye^{iK_{Ax}a}-A'^1_ye^{-iK_{Ax}a})=K_{Bx}(B^1_ye^{iK_{Bx}a}-B'^1_ye^{-iK_{Bx}a})
    \end{array}
   \right..
\end{eqnarray}
By Eq. (36), we can obtain the matrix form of $B^1_{y}$,
$B'^1_{y}$ and $A^1_{y}$, $A'^1_{y}$
\begin{eqnarray}
\left ( \begin{array}{ll}
B^1_{y}\\
B'^1_{y}\\
   \end{array}
\right )&&=\frac{1}{2} \left ( \begin{array}{ll}
(1+{K_{Ax}}/{K_{Bx}})e^{i(K_{Ax}-K_{Bx})a} \hspace{0.2in} (1-K_{Ax}/K_{Bx})e^{-i(K_{Bx}+K_{Ax})a}\\
(1-{K_{Ax}}/{K_{Bx}})e^{i(K_{Ax}+K_{Bx})a} \hspace{0.2in}
(1+{K_{Ax}}/{K_{Bx}})e^{i(K_{Bx}-K_{Ax})a}
   \end{array}
   \right )\left ( \begin{array}{ll}
A^1_{y}\\
A'^1_{y}\\
   \end{array}
\right )\nonumber \\&&=M_B^1\left ( \begin{array}{ll}
A^1_{y}\\
A'^1_{y}\\
   \end{array}
\right ),
\end{eqnarray}
where
\begin{eqnarray}
M_B^1=\frac{1}{2} \left ( \begin{array}{ll}
(1+{K_{Ax}}/{K_{Bx}})e^{i(K_{Ax}-K_{Bx})a} \hspace{0.2in} (1-K_{Ax}/K_{Bx})e^{-i(K_{Bx}+K_{Ax})a}\\
(1-{K_{Ax}}/{K_{Bx}})e^{i(K_{Ax}+K_{Bx})a} \hspace{0.2in}
(1+{K_{Ax}}/{K_{Bx}})e^{i(K_{Bx}-K_{Ax})a}
   \end{array}
   \right ),
\end{eqnarray}
is the transfer matrix of medium $B$ in the first period, and the
$K_{Bx}$ is
\begin{eqnarray}
K_{Bx}=\sqrt{K_{B}^{2}-K^{2}\sin^{2}\theta},
\end{eqnarray}
where $K_{B}=\frac{\omega}{c}n_b$, $n_b$ is the refractive indexes
of medium $B$.

Secondly, we use the similar approach can obtain the transfer
matrices $M_A^2$ and $M_B^2$ of media $A$ and $B$ in the second
period, they are
\begin{eqnarray}
M_A^2=\frac{1}{2} \left ( \begin{array}{ll}
(1+{K_B}/{K_A})e^{i(K_B-K_A)(a+b)} \hspace{0.1in}(1-{K_B}/{K_A})e^{-i(K_A+K_B)(a+b)}\\
(1-{K_B}/{K_A})e^{i(K_A+K_B)(a+b)} \hspace{0.1in}(1+{K_B}/{K_A})e^{i(K_A-K_B)(a+b)}\\
   \end{array}
   \right ),
\end{eqnarray}
and
\begin{eqnarray}
M_B^2=\frac{1}{2} \left ( \begin{array}{ll}
(1+{K_A}/{K_B})e^{i(K_A-K_B)(2a+b)} \hspace{0.2in}(1-{K_A}/{K_B})e^{-i(K_A+K_B)(2a+b)}\\
(1-{K_A}/{K_B})e^{i(K_A+K_B)(2a+b)} \hspace{0.2in}(1+{K_A}/{K_B})e^{i(K_B-K_A)(2a+b)}\\
   \end{array}
   \right ),
\end{eqnarray}
Finally, we can give the transfer matrices $M_A^N$ and $M_B^N$ of
media $A$ and $B$ in the N-th period, they are
\begin{eqnarray}
M_A^N=\frac{1}{2} \left ( \begin{array}{ll}
(1+{K_B}/{K_A})e^{i(K_B-K_A)(N-1)(a+b)} \hspace{0.1in}(1-{K_B}/{K_A})e^{-i(K_A+K_B)(N-1)(a+b)}\\
(1-{K_B}/{K_A})e^{i(K_A+K_B)(N-1)(a+b)} \hspace{0.1in}(1+{K_B}/{K_A})e^{i(K_A-K_B)(N-1)(a+b)}\\
   \end{array}
   \right ),
 \end{eqnarray}
and
\begin{eqnarray}
M_B^N=\frac{1}{2} \left ( \begin{array}{ll}
(1+{K_A}/{K_B})e^{i(K_A-K_B)(N(a+b)-b)} \hspace{0.2in}(1-{K_A}/{K_B})e^{-i(K_A+K_B)(N(a+b)-b)}\\
(1-{K_A}/{K_B})e^{i(K_A+K_B)(N(a+b)-b)} \hspace{0.2in}(1+{K_A}/{K_B})e^{i(K_B-K_A)(N(a+b)-b)} \\
   \end{array}
   \right ).
\end{eqnarray}
With the transform matrices, we can give their relations: \\
(a) The representation of the first period transform matrices are
\begin{eqnarray} \left ( \begin{array}{ll}
 A^1_{y}\\
 A'^1_{y}\\
   \end{array}
   \right )=M_A^1\left ( \begin{array}{ll}
F_{y}\\
F'_{y}\\
   \end{array}
   \right ),
\end{eqnarray}
\begin{eqnarray}
\left ( \begin{array}{ll}
B^1_{y}\\
 B'^1_{y}\\
   \end{array}
   \right )=M_B^1\left ( \begin{array}{ll}
A^1_{y}\\
A'^1_y\\
   \end{array}
   \right )=M_B^1M_A^1\left ( \begin{array}{ll}
F_y\\
F'_y\\
   \end{array}
   \right )=M^1\left ( \begin{array}{ll}
F_y\\
F'_y\\
   \end{array}
   \right ).
\end{eqnarray}\\

(b) The representation of the second period transform matrices are
\begin{eqnarray}
\left ( \begin{array}{ll}
A^2_{y}\\
 A'^2_{y}\\
   \end{array}
   \right )=M_A^2\left ( \begin{array}{ll}
B^1_{y}\\
B'^1_{y}\\
   \end{array}
   \right )=M_A^2M_B^1M_A^1\left ( \begin{array}{ll}
F_{y}\\
F'_{y}\\
   \end{array}
   \right )=M_A^2M^1\left ( \begin{array}{ll}
F_{y}\\
F'_{y}\\
   \end{array}
   \right ),
\end{eqnarray}
\begin{eqnarray}
\left ( \begin{array}{ll}
B^2_{y}\\
 B'^2_{y}\\
   \end{array}
   \right )=M_B^2\left ( \begin{array}{ll}
A^2_{y}\\
A'^2_{y}\\
   \end{array}
   \right )=M_B^2M_A^2M_B^1M_A^1\left ( \begin{array}{ll}
F_{y}\\
F'_{y}\\
   \end{array}
   \right )=M^2M^1\left ( \begin{array}{ll}
F_{y}\\
F'_{y}\\
   \end{array}
   \right ).
\end{eqnarray}\\

(c) Similarly, the representation of the N-th period transform
matrices are
\begin{eqnarray}
\left ( \begin{array}{ll}
A^N_{y}\\
 A'^N_{y}\\
   \end{array}
   \right )=M_A^NM_B^{N-1}M_A^{N-1}\cdot\cdot\cdot M_A^2M_B^1M_A^1\left ( \begin{array}{ll}
F_{y}\\
F'_{y}\\
   \end{array}
   \right )=M_A^NM^{N-1}\cdot\cdot\cdot M^2M^1\left ( \begin{array}{ll}
F_{y}\\
F'_{y}\\
   \end{array}
   \right ),
\end{eqnarray}
\begin{eqnarray}
\left ( \begin{array}{ll}
B^N_{y}\\
 B'^N_{y}\\
   \end{array}
   \right )=M_B^NM_A^NM_B^{N-1}M_A^{N-1}\cdot\cdot\cdot M_A^2M_B^1M_A^1\left ( \begin{array}{ll}
F_{y}\\
F'_{y}\\
   \end{array}
   \right )=M^NM^{N-1}\cdot\cdot\cdot M^2M^1\left ( \begin{array}{ll}
F_{y}\\
F'_{y}\\
   \end{array}
   \right )=M\left ( \begin{array}{ll}
F_{y}\\
F'_{y}\\
   \end{array}
   \right ),
\end{eqnarray}
where
\begin{eqnarray}
M=M^NM^{N-1}\cdot\cdot\cdot M^2M^1=\left ( \begin{array}{ll}
  m_1 \hspace{0.1in} m_2\\
m_3 \hspace{0.1in} m_4\\
   \end{array}
   \right ),
\end{eqnarray}
is the total transform matrix of N period, and $M^1=M_B^1M_A^1$ is
the first period transform matrix, $M^2=M_B^2M_A^2$ is the second
period transform matrix, and $M^N=M_B^NM_A^N$ is the N-th period
transform matrix.

The wave function of N-th period in medium $B$ is
\begin{eqnarray}
\vec{\psi}_{B}^{N}(x,y,z)&=&(B_{x}^{N}e^{i(K_{Bx}x+K_{Bz}z)}+B_{x}^{'N}e^{i(-K_{Bx}x+K_{Bz}z)})\vec{i}\nonumber\\
&+&(B_{y}^{N}e^{i(K_{Bx}x+K_{Bz}z)}+B_{y}^{'N}e^{i(-K_{Bx}x+K_{Bz}z)})\vec{j}\nonumber\\
&+&(B_{z}^{N}e^{i(K_{Bx}x+K_{Bz}z)}+B_{z}^{'N}e^{i(-K_{Bx}x+K_{Bz}z)})\vec{k},
\end{eqnarray}

In FIG. 1, the transmission wave function is
\begin{eqnarray}
\vec{\psi}_{D}(x,y,z)=D_xe^{i(K_xx+K_Zz)}\vec{i}+D_ye^{i(K_xx+K_Zz)}\vec{j}+D_ze^{i(K_xx+K_zz)}\vec{k}.
\end{eqnarray}
(4) At $x=N(a+b)$, by the $\vec{j}$ component continuation of wave
functions $\vec{\psi}_{B}^{N}(x,y,z)$ and $\vec{\psi}_{D}(x,y,z)$,
we have
\begin{eqnarray}
B_{y}^{N}e^{i(K_{Bx}N(a+b)+K_{Bz}z)}+B_{y}^{'N}e^{i(-K_{Bx}N(a+b)+K_{Bz}z)}=D_ye^{i(K_xN(a+b)+K_Zz)},
\end{eqnarray}
since the Eq. (53) is an equation for an arbitrary variable $z$,
we have
\begin{eqnarray}
B_{y}^{N}e^{iK_{Bx}N(a+b)}+B_{y}^{'N}e^{-iK_{Bx}N(a+b)}=D_ye^{iK_xN(a+b)},
\end{eqnarray}
with Eqs. (49) and (50), the Eq. (54) can be written as
\begin{eqnarray}
(m_1F_y+m_2F'_y)e^{iK_{Bx}N(a+b)}+(m_3F_y+m_4F'_y)e^{-iK_{Bx}N(a+b)}=D_ye^{iK_xN(a+b)},
\end{eqnarray}
(5) At $x=N(a+b)$, by the $\vec{j}$ component derivative
continuation of wave functions $\vec{\psi}_{B}^{N}(x,y,z)$ and
$\vec{\psi}_{D}(x,y,z)$, we have
\begin{eqnarray}
K_{Bx}B_{y}^{N}e^{iK_{Bx}N(a+b)}-K_{Bx}B_{y}^{'N}e^{-iK_{Bx}N(a+b)}=K_xD_ye^{iK_xN(a+b)},
\end{eqnarray}
with Eqs. (49) and (50), the Eq. (56) can be written as
\begin{eqnarray}
\frac{K_Bx}{K_x}(m_1F_y+m_2F'_y)e^{iK_{Bx}N(a+b)}-\frac{K_Bx}{K_x}(m_3F_y+m_4F'_y)e^{-iK_{Bx}N(a+b)}=D_ye^{iK_xN(a+b)}.
\end{eqnarray}
By Eqs. (55) and (57), we can obtain
\begin{eqnarray}
\frac{F'_y}{F_y}=\frac{m_1(K_x-K_{Bx})e^{iK_{Bx}N(a+b)}+m_3(K_x+K_{Bx})e^{-iK_{Bx}N(a+b)}}{m_2(K_{Bx}-K_x)e^{iK_{Bx}N(a+b)}-m_4(K_x+K_{Bx})e^{-iK_{Bx}N(a+b)}},
\end{eqnarray}
\begin{eqnarray}
t=\frac{D_y}{F_y}=(m_1+m_2\frac{F'_y}{F_y})e^{i(K_{Bx}-K_x)N(a+b)}+(m_3+m_4\frac{F'_y}{F_y})e^{-i(K_{Bx}+K_x)N(a+b)},
\end{eqnarray}
and the quantum transmissivity $T$ is
\begin{eqnarray}
T=|t|^2.
\end{eqnarray}

\begin{figure}[tbp]
\includegraphics[width=12 cm]{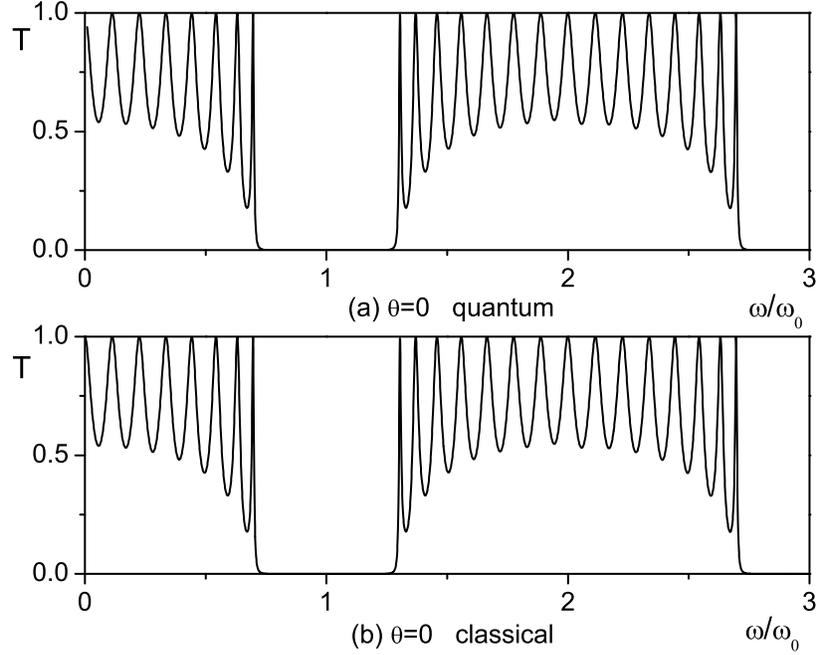}
\caption{The quantum and classical transmissivity of incident
angle $\theta=0$. (a) quantum transmissivity (b) classical
transmissivity. }
\end{figure}

\begin{figure}[tbp]
\includegraphics[width=12 cm]{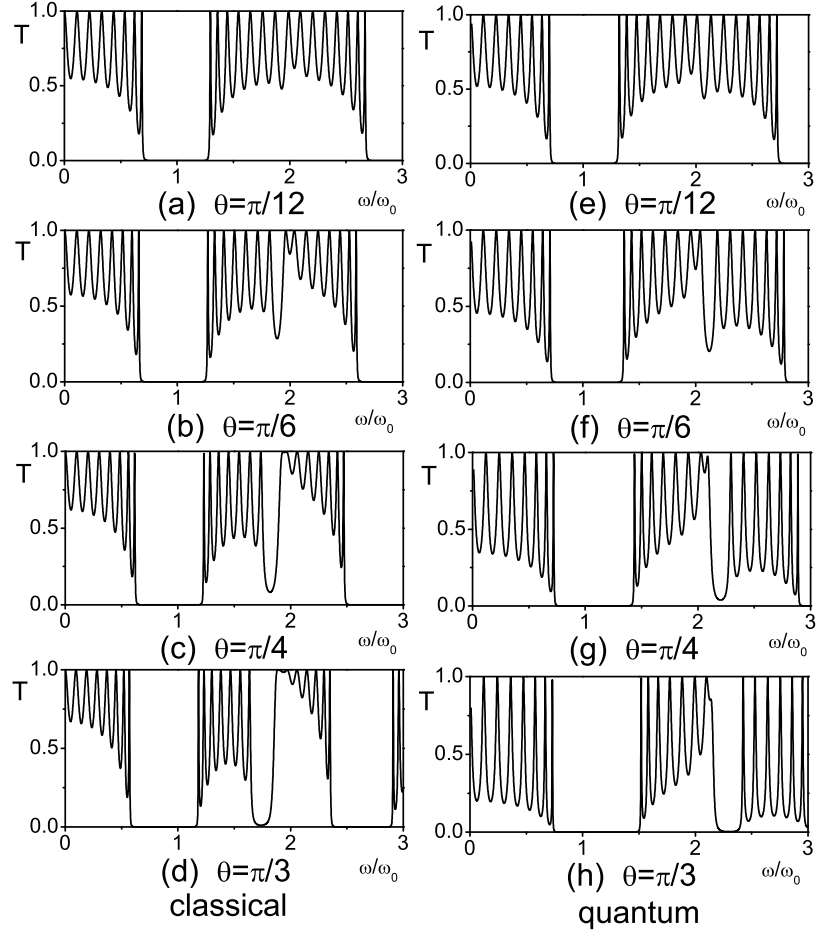}
\caption{The quantum and classical transmissivity for different
incident angle $\theta$. (a)-(d) classical transmissivity, (e)-(h)
quantum transmissivity. }
\end{figure}

\begin{figure}[tbp]
\includegraphics[width=12 cm]{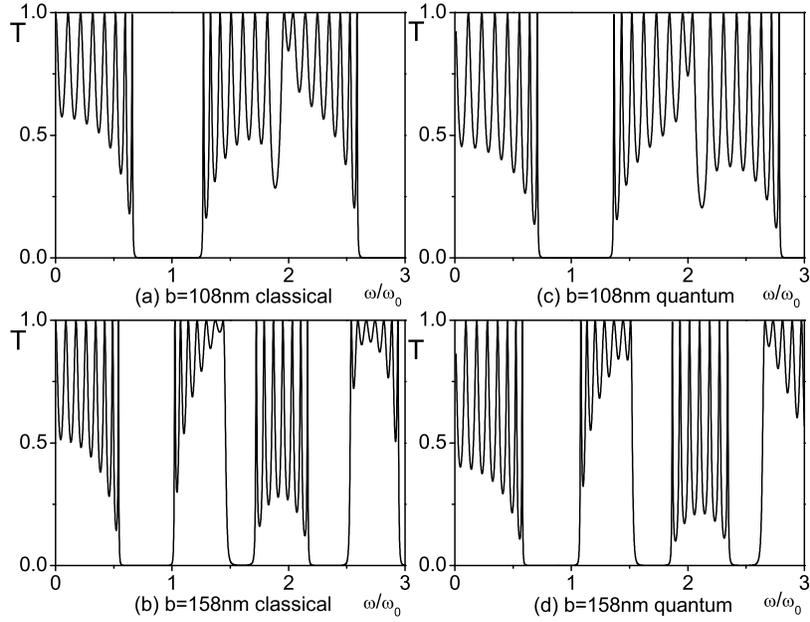}
\caption{The quantum and classical transmissivity of incident
angle $\theta=\frac{\pi}{6}$ for different thickness $b$. (a)-(b)
classical transmissivity, (c)-(d) quantum transmissivity. }
\end{figure}
\begin{figure}[tbp]
\includegraphics[width=12 cm]{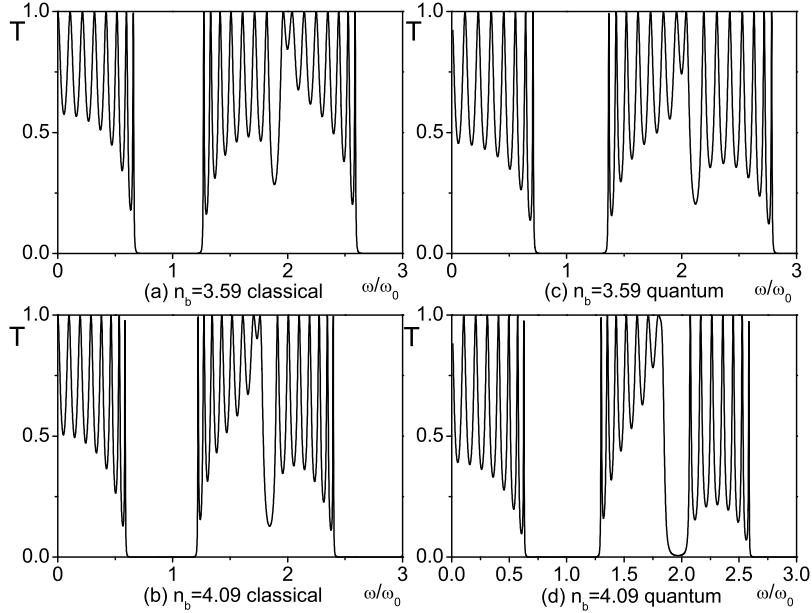}
\caption{The quantum and classical transmissivity of incident
angle $\theta=\frac{\pi}{6}$ for different refractive indexes
$n_b$. (a)-(b) classical transmissivity, (c)-(d) quantum
transmissivity.}
\end{figure}

\begin{figure}[tbp]
\includegraphics[width=12 cm]{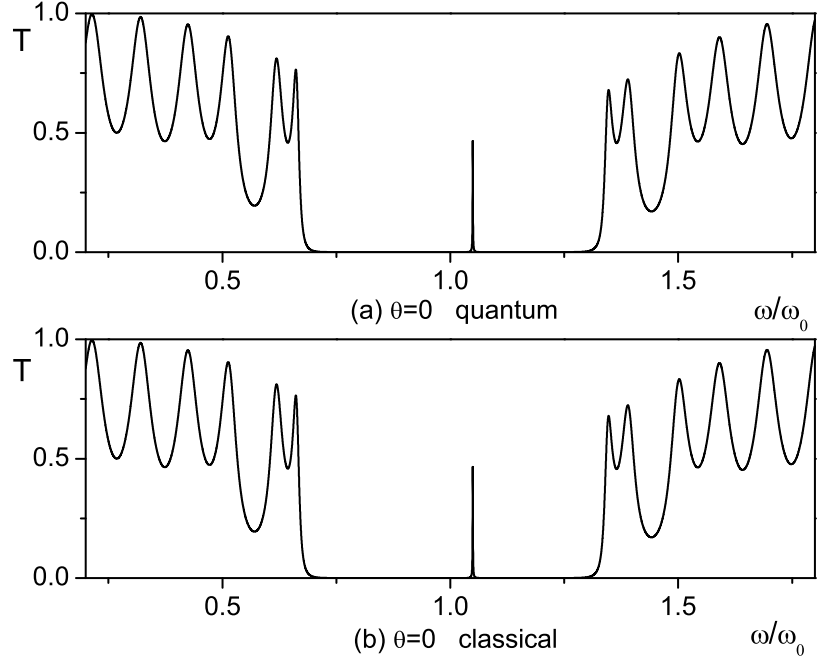}
\caption{The quantum and classical transmissivity of incident
angle $\theta=0$ with defect layer, the structure is
$(AB)^8D(AB)^8$. (a) quantum transmissivity (b) classical
transmissivity.}
\end{figure}

\begin{figure}[tbp]
\includegraphics[width=12 cm]{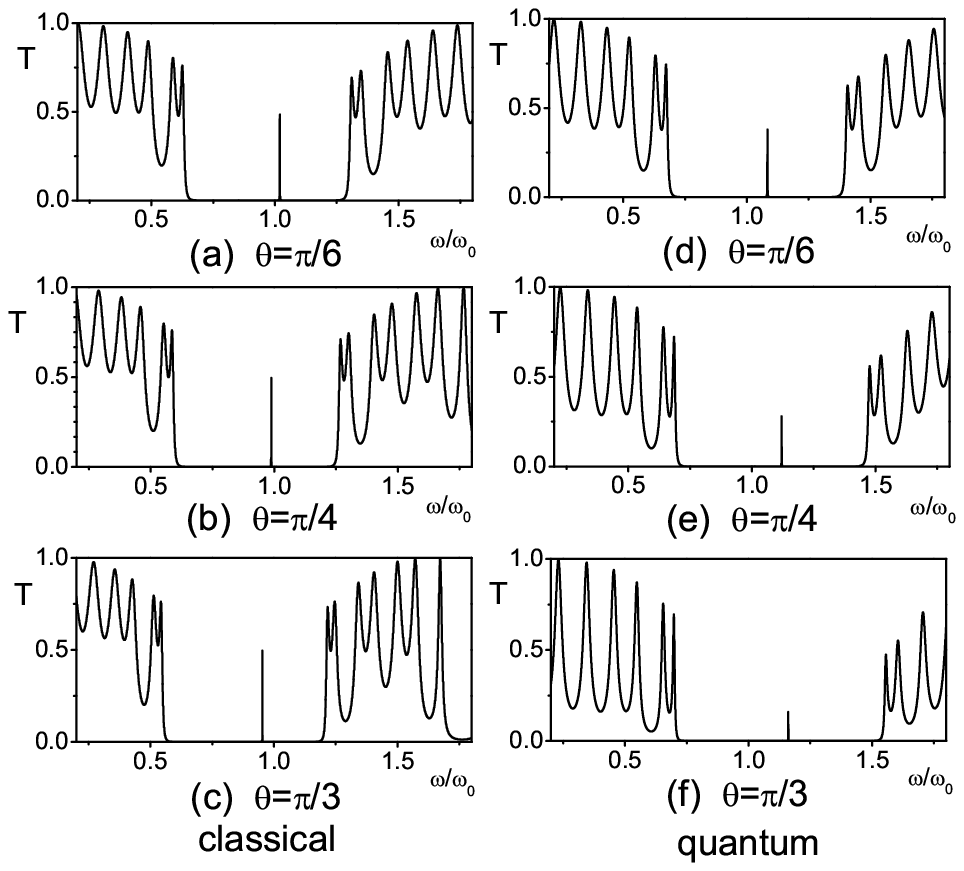}
\caption{The quantum and classical transmissivity for different
incident angle $\theta$ with defect layer, the structure is
$(AB)^8D(AB)^8$. (a)-(c) classical transmissivity, (d)-(f) quantum
transmissivity. }
\end{figure}

\vskip 5pt {\bf 5. Numerical result} \vskip 5pt

In this section, we report our numerical results of quantum
transmissivity. The main parameters are: For the medium $A$, its
refractive indexes is $n_a=1.45$, and thickness is $a=267nm$. For
the medium $B$, its refractive indexes is $n_b=3.59$, and
thickness is $b=108nm$. The central frequency is
$\omega_0=1.216\times10^{15}Hz$, and the period number is $N=16$.
With Eqs. (58)-(60), we can calculate the quantum transmissivity.
In FIG. 2, we calculate the quantum and classical transmissivity
when the incident angle $\theta=0$, FIG. 2 (a) and (b) are quantum
and classical transmissivity, respectively. Comparing FIG. 2 (a)
and (b), we find the quantum transmissivity and classical
transmissivity are the completely same, i.e., when the incident
angle $\theta=0$, there is not quantum effect in one-dimensional
photonic crystal. In FIG. 3, we calculate the quantum and
classical transmissivity when the incident angle $\theta$ are
$\frac{\pi}{12}$, $\frac{\pi}{6}$, $\frac{\pi}{4}$,
$\frac{\pi}{3}$, respectively. From FIG. 3 (a) to (d), they are
classical transmissivity. From FIG. 3 (e) to (h), they are quantum
transmissivity. We can obtain the following results: (1) With the
incident angle increase, the forbidden bands width are unchanged
and positions red shift for the classical transmissivity. (2) With
the incident angle increase, the forbidden bands become widened
and positions blue shift for the quantum transmissivity. (3) For
the same incident angle $\theta$, the quantum forbidden bands are
wider than the classical forbidden bands, and the quantum
forbidden bands positions blue shift. (4) When the incident angle
increase, the quantum effect become more remarkable. In FIG. 4, at
the incident angle $\theta=\frac{\pi}{6}$, we calculate the
quantum and classical transmissivity when the thickness $b$ of
medium $B$ are $108 nm$ and $158 nm$, respectively. FIG. 4 (a) and
(b) are the classical transmissivity, and FIG. 4 (c) and (d) are
the quantum transmissivity. We can obtain the following results:
(1) With the thickness $b$ increase, the forbidden bands numbers
increase and positions red shift for the classical and quantum
transmissivity. (2) For the same thickness $b$, the quantum
forbidden bands are wider than the classical forbidden bands, and
the quantum forbidden bands positions blue shift relative to the
classical forbidden bands. In FIG. 5, at the incident angle
$\theta=\frac{\pi}{6}$, we calculate the quantum and classical
transmissivity when the refractive indexes $n_b$ of medium $B$ are
$3.59$ and $4.09$, respectively. FIG. 5 (a) and (b) are the
classical transmissivity, and FIG. 5 (c) and (d) are the quantum
transmissivity. We can obtain the following results: (1) With the
refractive indexes $n_b$ increase, the forbidden positions red
shift for the classical and quantum transmissivity. (2) For the
same refractive indexes $n_b$, the quantum forbidden bands are
wider than the classical forbidden bands, and the quantum
forbidden bands positions blue shift relative to the classical
forbidden bands. In FIG. 6, we calculate the quantum and classical
transmissivity with defect layer, and the incident angle
$\theta=0$, FIG. 6 (a) and (b) are quantum and classical
transmissivity, respectively. Comparing FIG. 6 (a) and (b), we
find the classical and quantum transmissivity are identical, i.e.,
when the incident angle $\theta=0$, there is not quantum effect in
one-dimensional photonic crystal with defect layer. In FIG. 7, we
calculate the classical and quantum  transmissivity with defect
layer, and the incident angle are $\frac{\pi}{6}$,
$\frac{\pi}{4}$, $\frac{\pi}{3}$, respectively, and the structure
is $(AB)^8D(AB)^8$. From FIG. 7 (a) to (c), they are classical
transmissivity. From FIG. 7 (d) to (f), they are quantum
transmissivity. We can obtain the following results: (1) With the
incident angle increase, the forbidden bands width and defect
model intensity are unchanged, but the forbidden bands and defect
model positions red shift for the classical transmissivity. (2)
With the incident angle increase, the quantum forbidden bands
become widened, the defect model intensity weaken, and the
forbidden bands and defect model positions blue shift relative to
the classical transmissivity. (3) For the same incident angle, the
quantum forbidden bands are wider and defect model intensity
weaker than the classical forbidden bands, the defect model and
the quantum forbidden bands positions blue shift relative to the
classical. (4) When the incident angle increase, the quantum
effect become more remarkable.

\newpage
 \vskip 5pt
{\bf 6. Conclusion} \vskip 5pt

In summary, we have studied the quantum transmission
characteristics of one-dimensional photonic crystal by the quantum
theory approach, and compared the calculation results of classical
with quantum theory. We have found some quantum effects in
one-dimensional photonic crystal. When the incident angle
$\theta=0$, i.e., vertical incidence, the classical and quantum
transmission characteristics are the completely same, i.e., there
is not quantum effect in one-dimensional photonic crystal. When
the incident angle $\theta\neq0$, we find there are obvious
quantum effect with the incident angle increase. At the incident
angle $\theta\neq0$, there are also obvious quantum effect with
the change of thickness and refractive indexes of medium $B$ or
$A$. Otherwise, we have considered the effect of defect layer on
the quantum transmission characteristics. When the incident angle
$\theta=0$, there is also not quantum effect, and when the
incident angle $\theta\neq0$, with the incident angle increase,
there are obvious quantum effect for the one-dimensional photonic
crystal with defect layer. The new quantum effect of photonic
crystal shall help us to design optical devices.

\vskip 5pt {\bf 7.  Acknowledgment} \vskip 12pt

This work is supported by Scientific and Technological Development
Foundation of Jilin Province, Grant Number: 20130101031JC. 

\vskip
8pt



\begin{thebibliography}{10}

\bibitem{s1}
E. Yablonovitch, Phys. Rev. Lett. {\bf 58}, 2059 (1987).

\bibitem{s2}
S. John, Phys. Rev. Lett. {\bf 58} 2486 (1987).

\bibitem{s3}
A. Lavrinenko, P.I. Borel, L.H. Frandsen, M. Thorhauge, A. Harpth,
M. Kristensen, T. Niemi, Opt. Express {\bf 12} 234 (2004).

\bibitem{s4}
J. Pu, Y. Yomogida, K. K. Liu, L. J. Li, Y. Iwasa, and T.
Takenobu, Nano Lett., {\bf 12} 4013, (2012).

\bibitem{s5}
S. Fan, P.R. Villeneuve, J.D. Joannopoulos, H.A. Haus, Phys. Rev.
Lett. {\bf 80} 960 (1998).

\bibitem{s6}
A. Ferreira, N. M. R. Peres, R. M. Ribeiro and T. Stauber, Phys.
Rev. B, {\bf 85} 115438, (2012).

\bibitem{s7}
S. Kim, I. Park, H. Lim, Proc. Design of photonic crystal
splitters/combiners SPIE {\bf 5597} 129 (2004).

\bibitem{s8}
N. M. R. Peres and Yu. V. Bludov, EPL, {\bf 101} 58002, (2013).

\bibitem{s9}
R. Martinez-Sala, J. Sancho, J. V. Sanchez, V. Gomez, J. Llinares
and F. Meseguer, nature {\bf 378}, 241 (1995).

\bibitem{s10}
T. Cai, R. Bose, G. S. Solomon, and E. Waks, Appl. Phys. Lett.
{\bf 102} 141118 (2013).

\bibitem{s11}
J. J. Joannopoulos, R. D. Meade, J. N. Winn, Photonic crystals:
molding the flow of light (Princeton University Press, New Jersey,
1995).

\bibitem{s12}
M. Minkov and V. Savona, Phys. Rev. B {\bf 88} 081303R (2013).

\bibitem{s13}
K. K. Yee, IEEE Trans. Antennas Propag. {\bf 14}, 302 (1966).

\bibitem{s14}
K. S. Choi, H. Deng, J. Laurat, and H. J. Kimble, Nature {\bf
452}, 67 (2006).

\bibitem{s15}
J. B. Pendry, Phys. Rev. Lett. {\bf 69}, 2772 (1992).

\bibitem{s16}
M. Jachura, M. Karpinski, C. Radzewicz, and K. Banaszek, Opt.
Express {\bf 22} 8624 (2014).

\bibitem{s17}
D. A. Miller, Nature Photonics {\bf 4} 3 (2010).

\bibitem{s18}
Kamal, A., Clarke, J. Devoret, M. H, Nature Physics, {\bf 7} 311
(2011).

\bibitem{s19}
Fan, L., Science, {\bf 335} 447 (2012).

\bibitem{s20}
Yu Z., Fan S., Nature Photon. {\bf 3} 91 (2009).


\bibitem{s21}
Xiang-Yao Wu, Xiao-Jing Liu, and Yi-Heng Wu, et. al., Int J Theor
Phys, {\bf 49}, 194 (2010).

\bibitem{s22}
B. J. Smith and M. G. Raymer£¬Photon wave functions [J]¡£New J.
Phys. 9, 414 (2007).

\bibitem{s23}
Xiang-Yao Wu, Ji Ma, Xiao-Jing Liu, Jing-Hai Yang, Hong Li, Si-Qi
Zhang, Hai-Xin Gao, Xin-Guo Yin, San Chen, Physica E 59, 174
(2014).

\bibitem{s24}
Xiang-Yao Wu, Ji Ma, Xiao-Jing Liu, Jing-Hai Yang, Hong Li, Si-Qi
Zhang, Hai-Xin Gao, Heng-Mei Li, Hong-Chun Yuan, Optics
Communications 321, 211 (2014).

\end{thebibliography}
\end{document}